\documentclass[aps,prb,twocolumn,groupedaddress,showpacs]{revtex4-1}
\usepackage{graphicx,color}
\usepackage[colorlinks=true,urlcolor=blue,linkcolor=blue,
            citecolor=blue]{hyperref}

\mathchardef\mhyphen="2D

\begin{document}

\title{Energetics of H$_2$ clusters from density functional and
  coupled cluster theories}

\author{J. R. Trail}
\email{jrt32@cam.ac.uk}
\affiliation{Theory of Condensed Matter Group, Cavendish Laboratory, J
  J Thomson Avenue, Cambridge CB3 0HE, United Kingdom}

\author{P. L\'opez R\'ios}
\affiliation{Theory of Condensed Matter Group, Cavendish Laboratory, J
  J Thomson Avenue, Cambridge CB3 0HE, United Kingdom}

\author{R. J. Needs}
\affiliation{Theory of Condensed Matter Group, Cavendish Laboratory, J
  J Thomson Avenue, Cambridge CB3 0HE, United Kingdom}

\date{\today}

\begin{abstract}
  We use coupled-cluster quantum chemical methods to calculate the
  energetics of molecular clusters cut out of periodic molecular
  hydrogen structures that model observed phases of solid hydrogen.
  The hydrogen structures are obtained from Kohn-Sham density
  functional theory (DFT) calculations at pressures of 150, 250 and
  350 GPa, which are within the pressure range in which phases II, III
  and IV are found to be stable.
  The calculated deviations in the DFT energies from the
  coupled-cluster data are reported for different functionals, and
  optimized functionals are generated which provide reduced errors.
  We give recommendations for semi-local and hybrid density
  functionals that are expected to accurately describe hydrogen at
  high pressures.
\end{abstract}


\maketitle


The phase diagram of solid hydrogen has been studied theoretically
using a number of approaches such as density functional theory (DFT)
\cite{Jones_2015}, diffusion quantum Monte Carlo (DMC)
\cite{Ceperley_Alder_1980, Foulkes_RMP_2001}, and path integral
molecular dynamics (PIMD) methods \cite{PIMD_Marx_1996,
PIMD_Kitamura_2000, PIMD_H_Chen_2013, PIMD_H_Li_2013}.
The hexagonal structure of phase I of solid hydrogen was obtained from
experiment long ago \cite{Silvera_1980, Mao_1994}.
Structures for solid hydrogen at high-pressures have been proposed
from the DFT-based \textit{ab initio} random structure searching
method \cite{silane_Pickard_2006, structure_search_Pickard_2011,
structure_search_Needs_2016}.
These structures provide models for phases III
\cite{III_H_Pickard_2007}, IV \cite{DFT_H_IV_Pickard_2012} and, with
less certainty, for phase II \cite{DFT_H_II_2009} of solid hydrogen
which are in reasonable agreement with the available vibrational data
from infra-red and Raman spectroscopy \cite{Dalladay-Simpson_2016,
Zha_experiment_phase_IV_2013, Howie_phase_IV_H_and_D_2012,
Loubeyre_H_2002, Zha_IR_360_GPa_2012,
Howie_Mixed_Molecular_and_Atomic_Phase_hydrogen_2012,
Eremets_Conductive_dense_hydrogen_2011,
Goncharov_vibron_H2_D2_200GPa_2011}.
A hexagonal structure has recently been proposed as a model for the
low-pressure range of phase III \cite{Monserrat_2016}, but it is not
considered in this work.
DMC has been used to benchmark DFT results for solid hydrogen at high
pressures \cite{H_He_McMahon_2012, Morales_2013,
QMC_bulk_H_Azadi_2014, Clay_2014, QMC_bulk_H_Drummond_2015},
although the expense of the DMC calculations means that only very
limited data could be generated.

Here we construct exchange-correlation functionals that can be
expected to provide an accurate description of the energetics of the
structures mentioned above within DFT.
Two accurate hybrid functionals are provided, denoted O$_1$ and
O$_1$-D3, for use without and with dispersion corrections,
respectively.
We also provide accurate generalized gradient approximation (GGA)
functionals (though less accurate than their hybrid counterparts),
denoted O$_3$ and O$_3$-D3, for use without and with dispersion
corrections, respectively.
DMC energies are computed using orbitals generated with one of the
functionals and compared with results from coupled cluster singles,
doubles, and perturbative triples [CCSD(T)] in order to further
assess the accuracy of the hydrogen functionals.

\section{Generation of H$_2$ cluster geometries}

We use small molecular hydrogen clusters as reference systems to
construct our DFT functionals. 
We extract cluster geometries from stable bulk DFT structures obtained
with the PBE semi-local density functional by taking spherical cuts
of each system containing up to 24 hydrogen atoms.
Total energy calculations for a variety of clusters of hydrogen
molecules are carried out at the level of CCSD(T) and DFT.
The coupled cluster and DFT calculations are performed using the
\textsc{MOLPRO} code \cite{molpro}.

The procedure for generating cluster geometries is designed to give
reasonably compact and symmetric structures and is as follows.
For a bulk structure, atomic pairs closer than $0.85$\ \AA\ are
classified as molecules.
The distance parameter used in this classification is chosen so
that all atoms are classified as a member of only one molecule.
The center of mass of each H$_2$ molecule and the midpoint between the
center of mass of pairs of molecules is then recorded.

For clusters with an even number of molecules, a sphere is centered at
each midpoint and expanded until it contains the required number of
hydrogen atoms, $N$.
For clusters with an odd number of molecules, a sphere is placed at
the center of mass of each H$_2$ molecule and expanded until it
contains $N$ hydrogen atoms.
This procedure is carried out for each midpoint/center-of-mass, and
we select the point about which the magnitude of the first moment
of the selected atomic positions is minimized.
This process preserves much of the symmetry of the underlying bulk
structure, and supplies compact clusters representing each structure.

Clusters are generated from the bulk structures of symmetries
$P2_1c$-24, $C2/c$-24, and $Pc$-48, which model, respectively, phases
II, III, and IV of hydrogen, while $Cmca$-12 and $Cmca$-4 are other
low-energy structures that have been found in structure searches.
(Note that in this notation the number of atoms in the primitive unit
cell is given following the space group designation.)
The five structures are calculated using DFT at pressures of 150, 250,
and 350 GPa, which covers the experimentally relevant high-pressure
regime.
Clusters containing 1--12 pairs of hydrogen atoms are generated from
each of these structures, providing a total of 180 clusters.

\section{Extrapolation and error assessment for CCSD(T) and DFT}

CCSD(T) calculations are performed to provide reference energies for
each cluster which accurately include many-body effects.
These reference energies are then compared with DFT energies for
several exchange-correlation functionals.
The cc-pV$n$Z Gaussian basis sets are used \cite{basis_dunning_1992},
together with extrapolation to the complete basis set (CBS) limit.
Note that $n={\rm D}, {\rm T}, {\rm Q}, 5, 6, \ldots$ specifies the
largest value of the angular momentum index $l$ in the basis set.

A variety of procedures are available in the literature to extrapolate
CCSD(T) results to the CBS limit \cite{cbslim_feller_2013}.
We employ a computationally expensive extrapolation procedure, which
we refer to as ``accurate extrapolation'', for small clusters
containing 2--8 hydrogen atoms.
These ``accurate'' total energies are then used to validate an
empirical, but computationally cheaper, extrapolation procedure
that is applied to all cluster sizes, which we refer to as
``efficient extrapolation''.

The CCSD(T) energy is given by
\begin{equation}
  E_{\rm CCSD(T)} = E_{\rm HF} + E_{\rm S} + E_{\rm D}
                  + E_{\rm (T)} \;,
\end{equation}
where $E_{\rm HF}$ is the Hartree-Fock (HF) energy, $E_{\rm S}$ is the
second-order contribution to the correlation energy due to double
excitations of opposite-spin electrons, $E_{\rm D}$ is the
second-order contribution to the correlation energy due to double
excitations of like-spin electrons, and $E_{\rm (T)}$ is the
perturbative triple-excitation correlation energy.
We extrapolate the CCSD(T) energies of the smaller clusters to the CBS
limit using analytic forms justified by the expected asymptotic
behavior of the components of the CCSD(T) energy
\cite{extrap_ranasinghe_2013},
\begin{eqnarray}
  \label{eq:acc_extrap1}
  E_{\rm HF}(n)  &=& E_{\rm HF}(\infty)
                   + a_0 e^{ -b_0 \sqrt{n} }  \;, \\
  \label{eq:acc_extrap2}
  E_{\rm S}(n)   &=& E_{\rm S}(\infty)
                   + a_1 \left( n + c_1 \right)^{-3}
                   + b_1 \left( n + c_1 \right)^{-5} \;, \\
  E_{\rm D}(n)   &=& E_{\rm D}(\infty)
                   + a_2 \left( n + c_2 \right)^{-5}
                   + b_2 \left( n + c_2 \right)^{-7} \;, \\
  E_{\rm (T)}(n) &=& E_{\rm (T)}(\infty)
                   + a_3 \left( n + c_3 \right)^{-3}
                   + b_3 \left( n + c_3 \right)^{-5} \,,\;\;\;\;
\end{eqnarray}
where the left-hand sides of the equations are the components of the
CCSD(T) energy evaluated with the cc-pV$n$Z basis set, $\{a_i, b_i,
c_i\}$ are fitting parameters describing the variation of the energy
components with $n$, and $E_{\rm HF}(\infty)$, $E_{\rm S}(\infty)$,
$E_{\rm D}(\infty)$, and $E_{\rm (T)}(\infty)$ are the extrapolated
energy components, which are treated as fitting parameters.
Values of the left-hand sides are collected from CCSD(T) calculations
using basis sets with $n={\rm T}, {\rm Q}, 5, 6$, and a least-squares
fit for each component then provides the estimated CBS total energy.

This extrapolation is expected to be accurate due to the analytically
justified forms for each component of the total energy, and because
including higher order terms in the correlation energy expressions
ensures that the variation of energy with basis set size is accurately
described for the smaller basis sets.
The absolute deviation $|E_{\rm CCSD(T)}(6) - E_{\rm CCSD(T)}(\infty)|$
is expected to severely overestimate the error in this extrapolation
process, hence the mean and maximum absolute deviations of $4.6$ and
$5.4$ meV/[H$_2$] for clusters of 2--8 hydrogen atoms reliably
validate this finite basis correction.

Next, we use these ``accurate'' total energies to validate a
computationally cheaper extrapolation method for application to the
full set of clusters.
The variation of the total energy with basis set is dominated by the
$E_{\rm S}(n)$ term, hence we take the lowest-order part of
$E_{\rm S}(n)$ to construct our estimated CBS limit, $E_{\rm est}$.
We also estimate an extrapolation error interval by using similar
forms with different powers to define upper and lower limits,
$E_{\rm u}$ and $E_{\rm l}$, with values of the powers chosen so
that the estimated range includes the ``accurate'' estimates for the
small clusters of 2--8 hydrogen atoms.

The resulting ``efficient extrapolation'' procedure takes the form
\begin{eqnarray}
  E_{\rm u}(n)   &=& E_{\rm u}(\infty)
                   + a_{\rm u}   \left(n+n_0\right)^{-\frac{5}{2}}
                   \;, \\
  E_{\rm est}(n) &=& E_{\rm est}(\infty)
                   + a_{\rm est} \left(n+n_0\right)^{-3}
                   \;, \\
  E_{\rm l}(n)   &=& E_{\rm l}(\infty)
                   + a_{\rm l}   \left(n+n_0\right)^{-\frac{7}{2}}
                   \;,
\end{eqnarray}
where $E_{\rm u}(\infty)$, $E_{\rm est}(\infty)$, $E_{\rm l}(\infty)$,
$a_{\rm u}$, $a_{\rm est}$ and $a_{\rm l}$ are fitting parameters,
and $n_0$ is fixed to the average value of $c_1$ in
Eq.\ (\ref{eq:acc_extrap2}), with $n_0=-0.838$.
%
%
The two free parameters in each equation are obtained for each cluster
from CCSD(T) energies calculated using the two largest available basis
sets, resulting in an estimated CBS limit given by
$E_{\rm est}(\infty)$ and an estimated error of $[ E_{\rm u}(\infty) -
E_{\rm l}(\infty) ]/2$.

\begin{figure}[htb!]
  \includegraphics{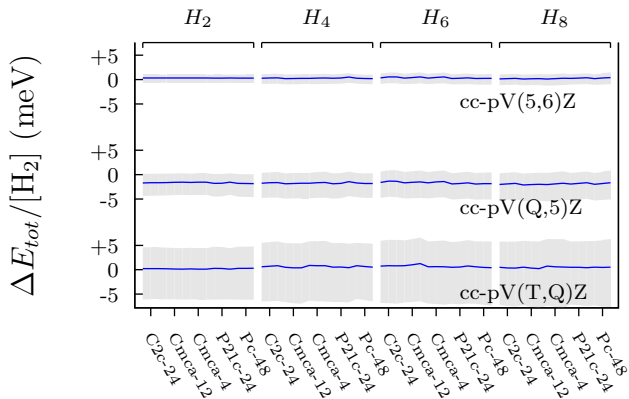}
  \caption{
    \label{fig:e_dev_from_accurate}
    ``Efficient'' estimates of CCSD(T) total energies, and the
    associated confidence intervals, plotted as the deviation from
    ``accurate'' total energies.
    Estimates are given for three different pairs of basis sets, and
    for geometries containing 8 hydrogen atoms or fewer.
    The gray shaded regions show the error ranges associated with the
    ``efficient extrapolation'' procedure.}
\end{figure}

For the subset of small clusters with $N=2$--8 the ``efficient''
procedure provides total energies that agree with the ``accurate''
values to within the estimated error ranges.
The mean absolute errors for small clusters using ``efficient'' (TQ),
(Q5) and (56) extrapolation are $6.1$, $2.6$, and $1.0$ meV/[H$_2$],
respectively, where we use the notation $(n_1 n_2)$ to denote
``efficient extrapolation'' from results with the cc-pV$n_1$Z and
cc-pV$n_2$Z basis sets.
The ``efficiently'' estimated energies and confidence intervals for
each small cluster geometry are shown in
Fig.~\ref{fig:e_dev_from_accurate} as a deviation from the
``accurate'' estimates.

The convergence of the DFT total energies with basis set is
exponential due to the correlation being contained in the
exchange-correlation functional rather than the wave function itself.
Consequently, extrapolation using only Eq.\ (\ref{eq:acc_extrap1})
provides accurate estimates of the CBS limit (replacing $E_{\rm HF}$
with the DFT energy).
In a similar manner to the CCSD(T) calculations, we begin with an
``accurate'' estimate of the finite basis correction for the
geometries with $N=2$--12.
The CBS total energy is provided by a least-squares fit of
Eq.\ (\ref{eq:acc_extrap1}) to results obtained with the $n={\rm D},
{\rm T}, {\rm Q}, 5$ basis sets, and with all parameters free.

The absolute deviation $|E_{\rm KS}(5) - E_{\rm KS}(\infty)|$ is
expected to severely overestimate the extrapolation error, hence mean
and maximum absolute deviations of $1.4$ and $4.4$ meV/[H$_2$] for
clusters of 2--12 hydrogen atoms reliably validate this finite-basis
correction.
Note that these mean and maximum deviations are taken over all of the
exchange-correlation functionals considered as well as the range of
cluster sizes.

Then, we construct an ``efficient'' estimate of the CBS limit by
two-point extrapolation using Eq.\ (\ref{eq:acc_extrap1}) with $b_0$
fixed to its average value for the ``accurate'' extrapolations,
where $b_0=6.707$.
Upper and lower limits are provided by extrapolation using the maximum
and minimum $b_0$ values ($b_0=8.247$ and $4.257$).
This provides estimated values and error ranges in a similar manner to
the CCSD(T) approach, with mean errors of $19.7$, $2.6$, and $0.9$
meV/[H$_2$] for (DT), (TQ), and (Q5) extrapolations, respectively.

Further basis set errors are expected to remain after extrapolation
due to superposition and the finite range of the cc-pV$n$Z basis sets.
These are found to be negligible (see Supplementary Material
\cite{supplementary}).

\section{Optimum exchange-correlation functionals}

Eight new exchange-correlation functionals are constructed by taking
the components of the B3LYP functional \cite{B3LYP_Kim_1994} and
determining their coefficients by minimizing the average difference
between the CCSD(T) and DFT energies using an iterative approach.
At iteration $i$, the total DFT energy provided by these B3LYP-like
functionals is of the form
\begin{eqnarray}
  E_{\rm KS}[\rho_i]
    &=& T[\rho_i] + V[\rho_i] +
        \alpha_0^{(i)} E_{\rm x}^{\rm HF}[\rho_i]  + \nonumber \\
    & & \alpha_1^{(i)} E_{\rm x}^{\rm B88}[\rho_i] +
        \alpha_2^{(i)} E_{\rm x}^{\rm LDA}[\rho_i] +           \\
    & & \alpha_3^{(i)} E_{\rm c}^{\rm LYP}[\rho_i] +
        \alpha_4^{(i)} E_{\rm c}^{\rm LDA}[\rho_i] +
                       E_{\rm D3} \;, \nonumber
\end{eqnarray}
where $\rho_i$ is the electronic density, $\alpha_j^{(i)}$ are linear
parameters, $E_{\rm x}^{\rm HF}$ is the exact Hartree-Fock exchange
energy, $E_{\rm x}^{\rm B88}$ and $E_{\rm c}^{\rm LYP}$ are the
BLYP-parameterized exchange and correlation functionals, and
$E_{\rm x}^{\rm LDA}$ and $E_{\rm c}^{\rm LDA}$ are the
parameterized local density approximation (LDA) exchange and
correlation functionals.
The final term is a dispersion correction, $E_{\rm D3}$.

The iterative method works as follows.
Parameter values at the start of iteration $i$, $\{\alpha_j^{(i)}\}$,
provide a DFT density $\rho_i$.
We then define the penalty function,
\begin{equation}
  P_i\left(\{\alpha_j^{(i)}\}\right) = \sum_k \left[
    \frac{  E^k_{\rm KS}[\rho_i]\left(\{\alpha_j^{(i)}\}\right)
          - E^k_{\rm CCSD(T)} }
         {N_k}
  \right]^2,
\end{equation}
where index $k$ runs over all 180 clusters considered, and
$E^k_{\rm CCSD(T)}$ is the best available ``efficiently extrapolated''
CCSD(T) energy for the $k$th cluster.
``Efficient'' (DT) extrapolation of DFT energies is used to limit the
computational cost.

The penalty function is then minimized with respect to
$\{\alpha_j^{(i)}\}$ with the density fixed to $\rho_i$.
Due to linearity this is easily achieved using matrix diagonalization,
and provides the optimum fixed-density parameters for $\rho_i$,
$\{\alpha_j^{(i+1)}\}$.
This process is then repeated until the variation in parameter values
and total energies is sufficiently small.

Four of the eight optimum functionals generated (O$_1$, O$_2$, O$_3$,
and O$_4$) do not include dispersion corrections ($E_{\rm D3}=0$),
while the rest (O$_1$-D3, O$_2$-D3, O$_3$-D3, and O$_4$-D3) include
that of Grimme \textit{et al.\@} \cite{disp_Grimme_2010}.
Dispersion corrections optimized for each functional are used when
available, and when unavailable BLYP corrections are used for GGA
functionals (LDA, PW91, O$_3$-D3 and O$_4$-D3) and B3LYP corrections
for hybrid functionals (SOGGA11-X, B3H, M08-SO, O$_1$-D3 and
O$_2$-D3).

We generate four hybrid functionals that include exact exchange
[O$_1$(-D3) and O$_2$(-D3)] and four GGA functionals that
exclude exact exchange [O$_3$(-D3) and O$_4$(-D3)].
Four of the functionals are constrained to obey the homogeneous
electron gas (HEG) limit [O$_2$(-D3) and O$_4$(-D3)], and the
remaining four [O$_1$(-D3) and O$_3$(-D3)] are not required to
obey the HEG limit.

Convergence is rapid for all eight functionals, with changes in the
total energy falling to less than $10^{-3}$ meV/[H$_2$] after four
iterations.
Optimum parameters and constraints are reported in Table
\ref{tab:optimum_parameters}.

The error in the total energy estimates used for optimization is
small.
The CBS total energies obtained with (DT) extrapolation incur a
maximum estimated error of 36 meV/[H$_2$].
This is an overestimate; the maximum difference between energies
obtained using (DT) extrapolation and those obtained using larger
basis sets (see the next section) is 8 and 13 meV/[H$_2$] for the
optimum functionals that satisfy or do not satisfy the HEG limit,
respectively.

\begin{table*}[htb!]
  \begin{tabular}{lr@{.}lr@{.}lr@{.}lr@{.}lr@{.}l}
    \hline \hline
    & \multicolumn{6}{c}{Exchange}
    & \multicolumn{4}{c}{Correlation} \\
    Functional
    & \multicolumn{2}{c}{$\alpha_0$}
    & \multicolumn{2}{c}{$\alpha_1$}
    & \multicolumn{2}{c}{$\alpha_2$}
    & \multicolumn{2}{c}{$\alpha_3$}
    & \multicolumn{2}{c}{$\alpha_4$} \\
    \hline
    O$_1$    & $ 0$&$60354910205605$
             & $ 0$&$14589659222789$
             & \multicolumn{2}{c}{($1-\alpha_0-\alpha_1$)}
             & $ 0$&$54819918079350$
             & \multicolumn{2}{c}{($1-\alpha_3$)} \\
    O$_2$    & $ 0$&$76679614121531$
             & $-0$&$84277505048128$
             & $ 1$&$05244207765216$
             & $-1$&$85005839047496$
             & $ 2$&$33023683529908$ \\
    O$_3$    & \multicolumn{2}{c}{($0$)}
             & $ 0$&$79872851641853$
             & \multicolumn{2}{c}{($1-\alpha_0-\alpha_1$)}
             & $ 0$&$61544187700225$
             & \multicolumn{2}{c}{($1-\alpha_3$)} \\
    O$_4$    & \multicolumn{2}{c}{($0$)}
             & $ 1$&$29167530141040$
             & $-0$&$25869459399816$
             & $ 2$&$03890740504165$
             & $-0$&$86103233660732$ \\[0.1cm]
    O$_1$-D3 & $ 0$&$49200298266553$
             & $ 0$&$34311594882336$
             & \multicolumn{2}{c}{($1-\alpha_0-\alpha_1$)}
             & $ 0$&$67727856886346$
             & \multicolumn{2}{c}{($1-\alpha_3$)} \\
    O$_2$-D3 & $ 0$&$61183767165887$
             & $-0$&$39108315837275$
             & $ 0$&$76213024756349$
             & $-1$&$10597328426091$
             & $ 1$&$72092162085949$ \\
    O$_3$-D3 & \multicolumn{2}{c}{($0$)}
             & $ 0$&$90916334640647$
             & \multicolumn{2}{c}{($1-\alpha_0-\alpha_1$)}
             & $ 0$&$78790801474948$
             & \multicolumn{2}{c}{($1-\alpha_3$)} \\
    O$_4$-D3 & \multicolumn{2}{c}{($0$)}
             & $ 1$&$38553958317866$
             & $-0$&$35884244275049$
             & $ 2$&$16733823510143$
             & $-0$&$96208190626903$ \\[0.1cm]
    B3LYP    & $ 0$&$20000000000000$
             & $ 0$&$72000000000000$
             & $ 0$&$08000000000000$
             & $ 0$&$81000000000000$
             & $ 0$&$19000000000000$ \\
    \hline \hline
  \end{tabular}
  \caption{
    \label{tab:optimum_parameters}
    Parameters that define the optimized functionals used.
    Parameters for the standard B3LYP functional are also shown,
    for comparison.}
\end{table*}

\section{Results}

We quantify the performance of exchange-correlation functionals by
their ability to reproduce the CCSD(T) total energies for the H$_2$
clusters.
In addition to our optimum functionals, we test 11 standard
functionals readily available in the literature: one LDA, four GGA,
and six hybrid functionals that combine exact exchange and semi-local
exchange-correlation functionals.
We test the functionals with and without dispersion corrections,
providing a total of 30 distinct DFT Hamiltonians (see Table
\ref{tab:functional_list}).

\begin{table}[hbt!]
  \begin{tabular}{ll}
    \hline \hline
    Functional                              & Type   \\
    \hline
    LDA                                     & Local  \\
    PBE        \cite{PBE_Perdew_1996}       & GGA    \\
    BLYP       \cite{BLYP_Becke_1988}       & GGA    \\
    revPBE     \cite{PBEREV_Zhang_1998}     & GGA    \\
    PW91       \cite{PW91_Perdew_1992}      & GGA    \\
    O$_3$(-D3)                              & GGA    \\
    O$_4$(-D3)                              & GGA    \\
    PBE0       \cite{PBE0_Adamo_1999}       & Hybrid \\
    SOGGA11-X  \cite{SOGGA11_Peverati_2011} & Hybrid \\
    B3H        \cite{B3H_chermette_1997}    & Hybrid \\
    B97        \cite{B97_Grimme_2006}       & Hybrid \\
    B3LYP      \cite{B3LYP_Kim_1994}        & Hybrid \\
    M08-SO     \cite{M08set_Zhao_2008}      & Hybrid (meta) \\
    O$_1$(-D3)                              & Hybrid \\
    O$_2$(-D3)                              & Hybrid \\
    \hline \hline
  \end{tabular}
  \caption{
    \label{tab:functional_list}
    Functionals used and their types.}
\end{table}

Calculations are performed using cc-pV$n$Z basis sets with different
$n$ for different clusters, and for CCSD(T) and DFT.
``Efficient'' extrapolation to the CBS limit, together with the
estimation of the extrapolation error, is performed as described
previously.

In the CCSD(T) calculations we use (56), (Q5), and (TQ) extrapolations
for clusters of $N=2$--$8$, $10$--$12$, and $14$--$24$ atoms,
respectively, obtaining mean basis set errors in the extrapolated
total energies of $1.0$, $2.9$, and $7.3$ meV/[H$_2$].
In the DFT calculations we use (Q5) and (TQ) extrapolations for
clusters of $N=2$--$12$ and $14$--$24$ atoms, respectively, obtaining
mean errors in the DFT energies of $0.9$ and $2.6$ meV/[H$_2$].
(For DFT mean values of errors are evaluated over all functionals, as
well as geometries.)
This energy resolution is sufficient to assess the accuracy of
exchange-correlation functionals.

\begin{figure}[htb!]
  \includegraphics{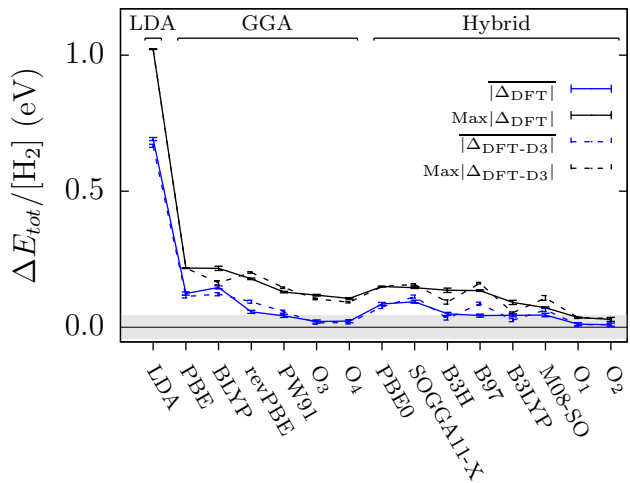}
  \caption{
    \label{fig:functionals_clusters}
    Maximum and mean absolute differences between DFT and CCSD(T)
    total energies for a variety of exchange-correlation functionals,
    both including ($\Delta_{\rm DFT{\mhyphen}D3}$) and excluding
    ($\Delta_{\rm DFT}$) dispersion corrections.
    The maximum and mean is taken over the 180 geometries considered,
    and basis set errors are almost imperceptible.
    Differences in the gray shaded region denote an accuracy of better
    than the chemical accuracy tolerance of $1$ kcal\,mol$^{-1}=43$
    meV/[H$_2$].}
\end{figure}

Each error is quantified as the deviation of the DFT energy from the
CCSD(T) energy, $\Delta = E_{\rm KS} - E_{\rm CCSD(T)}$.
Errors for each functional are summarized as both the mean and
maximum absolute difference for all clusters in
Fig.\ \ref{fig:functionals_clusters} for functionals
with and without dispersion corrections.

As expected, LDA energies are significantly worse than those from all
GGA and hybrid functionals, resulting in a peak error of ${\sim}1$
eV/[H$_2$] for H$_2$ that slowly falls with increasing cluster size.
Most of the LDA error may be identified with an imperfect correction
to the self-interaction (SI) included in the Coulomb potential for DFT
as this results in an error of $1.2$ eV/[H$_2$] for an isolated
hydrogen atom.
The improved accuracy of the non-LDA functionals is, in part, due to
their improved treatment of SI.

The general trend in Fig.\ \ref{fig:functionals_clusters} is for
hybrid functionals to be more accurate than GGA functionals, with the
optimized functionals providing the smallest errors for both types.
However, there is considerable overlap between errors for the two
functional types.
Of the hybrid functionals, only B3LYP and the optimized functionals
are more accurate than the optimized GGA functionals, whether or not
dispersion is present.

Given the fitting procedure and the extra variational freedom allowed
by relaxing the HEG limit, we would expect O$_2$-D3 and O$_4$-D3
functionals to reproduce the CCSD(T) results to the highest accuracy.
This is the case, with the lowest peak error for the hybrid
functionals of 32(3) meV/[H$_2$] provided by O$_2$-D3, and the lowest
peak error for the GGA functionals of 92(3) meV/[H$_2$] provided by
O$_4$-D3.
However, the O$_1$-D3 and O$_3$-D3 functionals, which are constrained
to reproduce the HEG limit, perform only marginally worse, with peak
errors of 35(3) meV/[H$_2$] and 104(3) meV/[H$_2$], respectively.


\begin{figure*}[htb!]
  \includegraphics{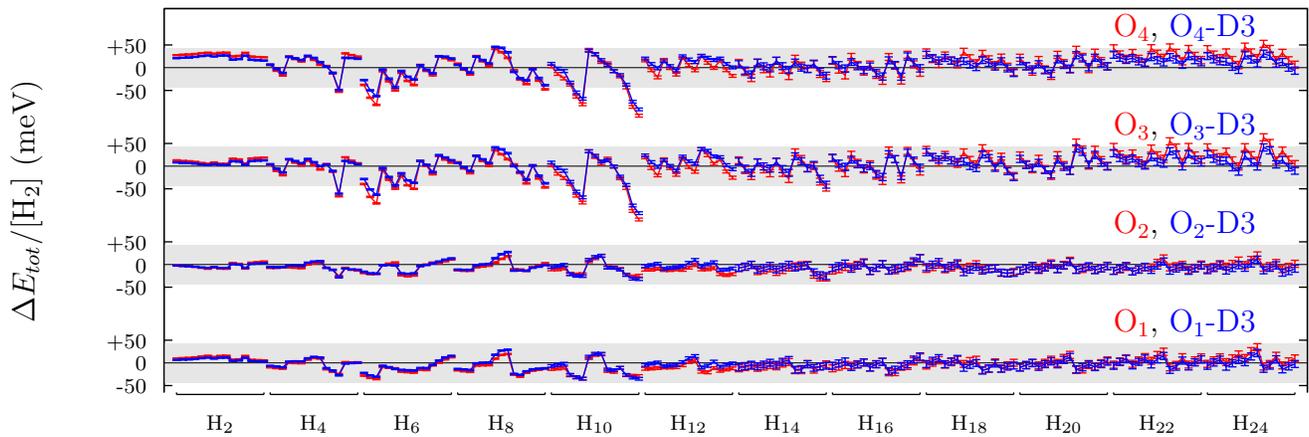}
  \caption{
    \label{fig:optimum_clusters}
    Deviation of DFT total energies from the CCSD(T) total energies.
    Results with (blue) and without (red) dispersion are shown.
    Values and errors bars are obtained using the ``efficient''
    estimates described in the text, and differences in the gray
    shaded regions denote an accuracy of better than the chemical
    accuracy tolerance of $1$ kcal\,mol$^{-1}=43$ meV/[H$_2$].}
\end{figure*}

Dispersion provides no consistent improvement to the performance of
the functionals considered.
The introduction of dispersion increases the peak error for 5 of the
15 functionals, leaves it unchanged for another 5, and decreases it
for the other 5 functionals.
Unsurprisingly, functionals optimized with dispersion give marginally
better results when applied with dispersion than without, and vice
versa.
This effect is not shown in the figures.

In order to gain further insight into the performance of the
functionals, in Fig.\ \ref{fig:optimum_clusters} we plot the deviation
of the DFT energies using the O$_n$(-D3) functionals from the CCSD(T)
energies for each of the 180 clusters.
Data generated with and without dispersion exhibit the same structure,
confirming that dispersion has a small effect on the results.
Overall, the hybrid functionals reproduce the CCSD(T) for all clusters
to within chemical accuracy, typically associated with a tolerance of
1 kcal\,mol$^{-1} = 43$ meV/[H$_2$].

For the optimum GGA functionals, O$_3$(-D3) and O$_4$(-D3), the
deviations from CCSD(T) are larger in magnitude, occasionally
exceeding the chemical accuracy tolerance, and display more structure
than those for the hybrid functionals.
Figure \ref{fig:optimum_clusters} shows segments of steep
(approximately) linear variation with the pressure of the underlying
bulk structure, where each segment is associated with a different
symmetry and cluster size.
This dependence is barely apparent for the hybrid functionals.

The inaccuracy of the functionals considered is most apparent
for cluster with 4--10 hydrogen atoms, and particularly for the
H$_{10}$ clusters.
This characteristic feature is identifiable for all of the GGA
functionals considered, and for most of the hybrid functionals [it is
not apparent for the B3H, M08-SO, O$_1$(-D3), and O$_2$(-D3)
functionals].

For a number of the functionals [though not the O$_n$(-D3)] there is
an underlying linear variation of error with cluster size.
This observation, together with the known inadequate description of SI
by DFT, suggests that the SI error should be included in our
assessment of the accuracy of the functionals.

We estimate the SI error as the difference between the total DFT
energy of an isolated hydrogen atom evaluated with each functional
(using an $n=6$ finite basis set that results in a
negligible basis set error) and the exact analytic energy.

\begin{figure}[htb!]
  \includegraphics{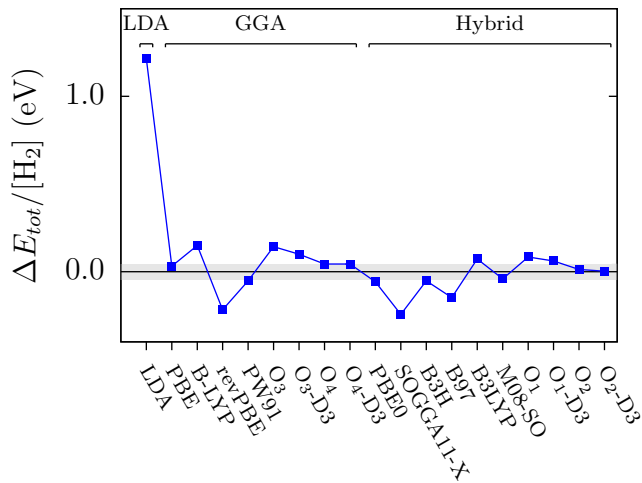}
  \caption{
    \label{fig:functionals_atom}
    Self-interaction errors for an isolated hydrogen atom for the 19
    functionals considered in the text, and using the exact density.
    Points within the gray shaded region denote a self-interaction
    error smaller than the chemical accuracy tolerance of 1
    kcal\,mol$^{-1}=43$ meV/[H$_2$].}
\end{figure}

Figure \ref{fig:functionals_atom} shows the resulting SI error for
each functional (dispersion is zero for an isolated atom, so there
are only 19 distinct total energies to consider).
The greatest SI error arises for the LDA functional.
For the optimum functionals the SI error is consistently decreased by
including Hartree-Fock exchange and not enforcing the HEG limit.
Of all the functionals considered the SI error is smallest for the
O$_2$ and O$_2$-D3 hybrid functionals, taking values of $13$ and $3$
meV/[H$_2$], respectively.


Since we are seeking an accurate functional for systems very unlike
the HEG, relaxing the requirement that the HEG limit is conserved in
the functionals is justified.
However, it is reasonable to expect that a physically realistic
functional will not radically misrepresent the HEG.
Whether this is so may be assessed by summing the exchange and
correlation coefficients in Table \ref{tab:optimum_parameters}.
For all of the optimized functionals the HEG exchange is well
preserved, with total HEG exchange deviating from the true value by
less than 4\%.
The HEG correlation is significantly less accurate when the HEG limit
is not enforced, deviating from the true value by more than 17\% for
the GGA functionals, and more than 38\% for the hybrid functionals.

The proposed functionals can be used to generate single-particle
orbitals for quantum Monte Carlo calculations \cite{Foulkes_RMP_2001},
which is particularly useful for studies of crystalline hydrogen
structures, such as those of Refs.\@
\onlinecite{QMC_bulk_H_Drummond_2015} and
\onlinecite{QMC_bulk_H_Azadi_2014}, which require more accuracy than
afforded by DFT and much larger systems sizes than are accessible with
CCSD(T).
In order to assess the accuracy of the functionals in this context
we perform diffusion Monte Carlo (DMC) calculations using the
\textsc{casino} code \cite{CASINO_review} for the H$_6$, H$_{12}$,
and H$_{24}$ clusters using single-particle orbitals generated with
the O$_3$ GGA functional.
The orbitals are cusp-corrected \cite{Ma_cusps_2005} to
prevent divergences of the local energy at electron-nucleus
coalescence points, and used in a Slater-Jastrow
\cite{Drummond_jastrow_2005, LopezRios_jastrow_2011} trial wave
function, whose parameters are optimized using linear least-squares
energy minimization \cite{Umrigar_emin_2007, Toulouse_emin_2007} at
the variational Monte Carlo level.
The DMC energies are obtained using the recent modifications of
Zen \textit{et al.\@} \cite{Zen_timestep_2016}, and extrapolated
to zero timestep \cite{Lee_strategies_2011} and infinite population
size; additional details are given in the Supplementary Material
\cite{supplementary}.

\begin{figure}[htb!]
  \includegraphics{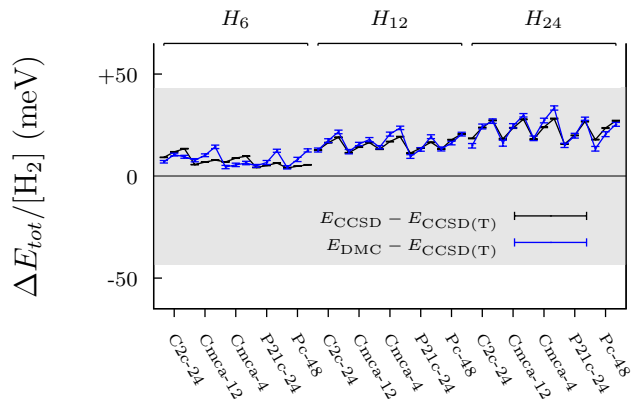}
  \caption{
    \label{fig:dmc_o3_e_dev}
    DMC and CCSD energies relative to CCSD(T) for the H$_6$, H$_{12}$,
    and H$_{24}$ clusters using single-particle orbitals obtained with
    the O$_3$ GGA functional.
    The error bars represent the statistical uncertainty in the DMC
    data, and the estimated basis-set extrapolation error in the
    CCSD data.
    All energy differences are smaller than the chemical accuracy
    tolerance of $1$ kcal\,mol$^{-1}=43$ meV/[H$_2$] indicated by the
    gray shaded region.}
\end{figure}

The DMC energies, shown in Fig.\ \ref{fig:dmc_o3_e_dev}, are higher
than the CCSD(T) energies for all 45 clusters.
Each of the successive S, D, and (T) contributions to the CCSD(T)
energy is about an order of magnitude smaller than the previous one,
which is expected in well-converged calculations, hence the missing
correlation energy in our CCSD(T) results may reasonably be identified
with the basis set error.
Figure \ref{fig:dmc_o3_e_dev} also shows close agreement between DMC
and CCSD energies, strongly suggesting that the small correlation
energy absent from our DMC results can be identified with the triple
excitations included in CCSD(T).

The energy differences CCSD(T) and DMC are smaller than $1$
kcal\,mol$^{-1}$, indicating that the single-particle orbitals
generated with the O$_3$ functional are an excellent choice for
performing accurate DMC calculations.
The pressure dependence of the DMC energies can be attributed to the
loss of accuracy of the DMC method at higher pressures.

In summary, the optimized functionals provide the best description of
the total energies for the 180 clusters considered, with optimized
hybrid functionals being the most accurate.
Relaxing the natural requirement that the functionals reproduce the
HEG limit provides an insignificant or small increase in the accuracy
of the functionals, at the cost of an inaccurate description of the
HEG correlation.
Dispersion corrections are small for the optimized functionals.

These points suggest that the functionals of choice for hydrogen
systems is the hybrid functional O$_1$(-D3), with O$_3$(-D3) a second
choice if a GGA is preferred.
The two different forms are available for application with or without
dispersion corrections.
With dispersion excluded, these provide an accuracy of better than
118(3) and 36(1) meV/[H$_2$] for GGA and hybrid functionals,
respectively.
With dispersion included, these provide an accuracy of better than
104(3) and 35(3) meV/[H$_2$] for GGA and hybrid functionals,
respectively.

There are several obvious options to improve the accuracy of optimized
functionals for hydrogen clusters.
The most evident improvement would be to reduce the remnant of basis
set error that arises from the extrapolation to the CBS limit used
during functional optimization.
This would be computationally expensive, and is probably not
significant given that the structure shown in
Fig.\ \ref{fig:functionals_atom} is accurately reproduced by (DT),
(TQ), or (Q5) extrapolation, so is not due to basis set error.
Another simple option would be to dynamically allocate weights in the
optimization process in order to equalize the distribution of errors
between clusters.
This would reduce the peak error at the cost of introducing a larger
error for clusters containing 12 hydrogen atoms or more.

Perhaps the most promising option for improving the functionals
would be to add more variational freedom to the optimized functional.
The simplest approach would be to include other parameterizations of
semi-local functionals, including meta terms, as this would preserve
the linearity of the optimization we rely on within the iterative
optimization process.
A more general approach would be to directly optimize the enhancement
factors present within the functional forms, introducing non-linearity
into each optimization iteration.
Similarly, the screening parameters present in the dispersion
correction may also be optimized, but it seems likely that this will
only provide a marginal improvement in the accuracy of the energies.

\begin{acknowledgments}
  The authors acknowledge financial support from the Engineering and
  Physical Sciences Research Council (EPSRC) of the
  U.K.\ [EP/J017639/1].
  Computational resources were provided by the University of Cambridge
  High Performance Computing Service
  (\url{http://www.hpc.cam.ac.uk}).
  Supporting research data may be freely accessed at [URL], in
  compliance with the applicable Open Data policies.
\end{acknowledgments}


\end{document}